\documentclass[twocolumn,showpacs,superscriptaddress,amsmath,amssymb]{revtex4}

\usepackage{graphicx}
\usepackage{dcolumn}
\usepackage{bm}
\usepackage{amstext}
\usepackage{color}
\draft \preprint{Submitted to Physical Review Letters}

\begin{document}
\title{Spin Chirality in a Molecular Dysprosium Triangle: the Archetype of the
Non-Collinear Ising Model
 }
\author{Javier Luzon}\affiliation{Laboratory of Molecular Magnetism, Department of
Chemistry and INSTM (UdR Firenze),
Universit\'{a} degli Studi di Firenze, Via della Lastruccia 3, 50019, Sesto Fiorentino,
Italy}
\author{Kevin Bernot}
\affiliation{Laboratory of Molecular Magnetism, Department of Chemistry and INSTM (UdR
Firenze),
Universit\'{a} degli Studi di Firenze, Via della Lastruccia 3, 50019, Sesto Fiorentino,
Italy}
\author{Ian J. Hewitt}
\affiliation{Institut f\"ur Anorganische Chemie, Universit\"at Karlsruhe (TH),
Engesserstr. 15, 76131
Karlsruhe, Germany}
\author{Christopher E. Anson}
\affiliation{Institut f\"ur Anorganische Chemie, Universit\"at Karlsruhe (TH),
Engesserstr. 15, 76131
Karlsruhe, Germany}
\author{Annie K. Powell}
\affiliation{Institut f\"ur Anorganische Chemie, Universit\"at Karlsruhe (TH),
Engesserstr. 15, 76131
Karlsruhe, Germany}
\author{Roberta Sessoli\footnote{corresponding author: roberta.sessoli@unifi.it}}
\affiliation{Laboratory of Molecular Magnetism, Department of Chemistry and INSTM (UdR
Firenze),
Universit\'{a} degli Studi di Firenze, Via della Lastruccia 3, 50019, Sesto Fiorentino,
Italy}

\date{\today}
\begin{abstract}

  Single crystal magnetic studies combined with a theoretical analysis show that
  cancellation of the magnetic moments in the trinuclear Dy$^{3+}$ cluster
  [Dy$_3$($\mu_3$-OH)$_2$L$_3$Cl(H$_2$O)$_5$]Cl$_3$, resulting in a non-magnetic ground
  doublet, originates from the non-collinearity of the single ion easy axes of
  magnetization of the Dy$^{3+}$ ions that lie in the plane of the triangle at 120$^{\circ}$ one from each other. This gives rise to a peculiar chiral nature of the ground
  non-magnetic doublet and to slow relaxation of the magnetization with abrupt accelerations at the crossings of the discrete energy levels.

\end{abstract}


\maketitle

\begin{figure}[!b]
\begin{center}
       \includegraphics[width=0.45\textwidth]{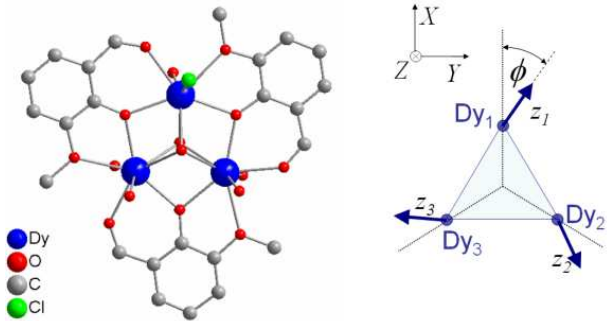}
\end{center}
\caption{(color online) Left:View of the molecular structure of Dy$_3$ cluster where the
hydrogen atoms, the chloride counteranions and the solvent molecules of crystallization
have been omitted. Right: Schematic view of the spin structure of the Dy$_3$ triangular
cluster and of the local easy axes orientation in respect of the laboratory $XYZ$
reference frame.}
\label{fig:triangle}
\end{figure}

Molecular nanomagnetism has provided benchmark systems to investigate new and
fascinating phenomena in magnetism\cite{GatteBook,Christou05} like magnetic memory at
the molecular level\cite{sessoli93}, quantum tunneling of the
magnetization\cite{Friedman96,Thomas96}, or destructive interferences in the tunneling
pathways\cite{Werns99}.	In this field rare earth ions like dysprosium(III) are currently
investigated because of their large magnetic anisotropy and high magnetic
moment\cite{Ishi03}. In the course of our synthetic efforts to obtain new molecular
nanomagnets based on rare-earth ions we recently obtained the trinuclear Dy$^{3+}$
cluster [Dy$_3$($\mu_3$-OH)$_2$L$_3$Cl(H$_2$O)$_5$]Cl$_3$ (where L is the anion of
ortho-vanillin)\cite{Tang06}, hence abbreviated as Dy$_3$, which possesses an almost
trigonal (C$_{3h}$) symmetry (see Figure \ref{fig:triangle} and EPAPS for more information)\cite{EPAPS}.

Preliminary powder magnetic measurements measurements on two compounds containing such Dy$_3$ clusters, but differing in their crystal packing and intermolecular contacts, revealed an identical magnetic behavior where the magnetization vs. field curve at low temperature is almost flat, suggesting a non magnetic ground state, but suddenly increases to its saturation value, $M_s$, at $H$ = 8 kOe\cite{Tang06}. This is in contrast to what commonly observed for antiferromagnetic triangular clusters that show a multi-step magnetization curve \cite{Luban02}. A possible explanation is that the Dy$^{3+}$ ions are characterized by almost ideal Ising anisotropy with the single ion easy axes lying in the plane of the triangle at 120$^{\circ}$ from one another, as represented in the scheme of Figure
\ref{fig:triangle}. To the best of our knowledge an experimental realization of this
simple but fascinating spin structure is unprecedented.

\begin{figure}[!h]
\begin{center}
       \includegraphics[width=0.38\textwidth]{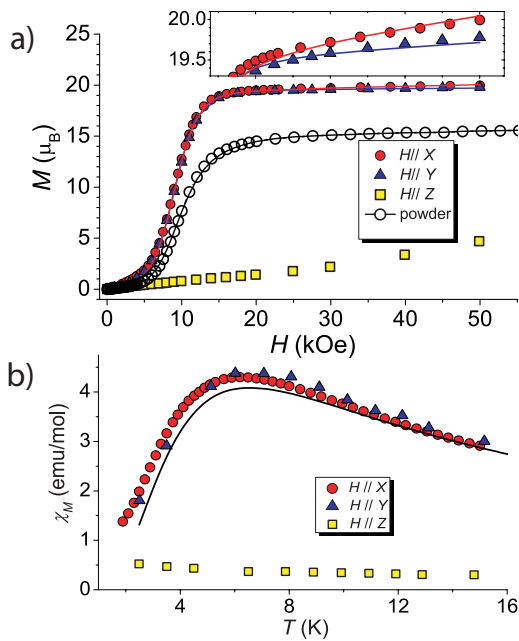}
\end{center}
\caption{(color online) a) Field dependence of the magnetization measured along the \emph{X},
 \emph{Y}, and  \emph{Z} axes at  \emph{T}=1.9 K. In the inset a magnified view of the high field region is
reported. b) Temperature dependence for the susceptibility along  \emph{X},  \emph{Y}, and  \emph{Z} axes with
an applied field of  1 kOe. The solid lines represent the calculated value with Eq. (2)
and the best fit parameters discussed in the text.
} \label{fig:squid}
\end{figure}

To verify our hypothesis larger crystals (of size \emph{ca.} 1 mm$^3$) of one of the two compounds were grown according to \cite{Tang06} through very slow evaporation of the solvent. This allowed an accurate face indexing of the crystal on the X-ray diffractometer and the investigation of the magnetic anisotropy by using an horizontal sample rotator in the SQUID magnetometer (see EPAPS for experimental details)\cite{EPAPS}. Scans in different crystallographic planes allowed us to determine the three magnetic anisotropy axes,
denoted as $X$, $Y$ and $Z$. The two structurally equivalent Dy$_3$ molecules in the unit cell
have the  Dy$_3$ planes almost perpendicular to  $Z$  and one side of the triangle
parallel to $Y$ (see Fig. \ref{fig:triangle}). Magnetization vs. field curves along these axes are given in Figure \ref{fig:squid}a. Along $X$ and $Y$ a sudden jump around 8 kOe is observed while an
almost linear but weaker magnetization is observed along $Z$. In Figure \ref{fig:squid}b
the temperature dependence of the magnetization along the three axes measured at 1 kOe, and
thus before the jump to saturation, are shown. The in-plane $X$ and $Y$ directions are very
similar and $M$ tends to zero at low temperatures, confirming a non-magnetic ground state,
while a weaker signal is observed along $Z$.

The observed behaviour has been modelled using the canonical formalism of the statistical thermodynamics and taking into account that Dy$^{3+}$ ions have a very large magnetic single-ion anisotropy due to the crystal field splitting
of the $^6H_{15/2}$ ground state. In a first approximation each Dy$^{3+}$ ion, which is supposed to have the doublet ground state well separated in energy from the other excited Stark sublevels, can be represented by an effective spin $S=1/2$ and the Dy$_3$
system can be modelled by an Ising Hamiltonian:

\begin{equation}
H=-\sum_{i,k=1,2,3}^{i>k}j_{zz}S_{z_i}S_{z_k}-\mu_B \sum_{i=1,2,3}g_zH_zS_{z_i}
\label{eq:hamilton1}
\end{equation}

where for each Dy$^{3+}$ ion the $z_i$ local axis is considered to be in the plane of
the triangle at 120$^{\circ}$  one from each other as schematized in Figure
\ref{fig:triangle}. The basis set of the full system consists therefore of 8 vectors:
$|\uparrow \uparrow \uparrow \rangle$, $|\downarrow \uparrow \uparrow \rangle$,
$|\uparrow \downarrow \uparrow \rangle$, $|\uparrow \uparrow \downarrow \rangle$,
$|\uparrow \downarrow \downarrow \rangle$ , $|\downarrow \uparrow \downarrow \rangle$,
$|\downarrow \downarrow \uparrow \rangle$, $|\downarrow \downarrow \downarrow \rangle$,
where, however, up and down refers to the local $z$ axes. Actually, these vectors are already the eigenvectors of (1).
	The best-fit parameters of the power data were $j_{zz}/k_B = 10.6(4) K$ and  $g_z =
20.7(1)$. This last value confirms that the ground doublet for the Dy$^{3+}$ ions is
well described by $|J=15/2, m_J=\pm15/2\rangle$. In fact  $J=15/2$ results from the
coupling of $L=5$ and $S =5/2$ and thus $g_J=4/3$, which,  for $m_J=\pm15/2$, gives an
effective gyromagnetic factor of $20$.

\begin{figure}[!t]
\begin{center}
       \includegraphics[width=0.37\textwidth]{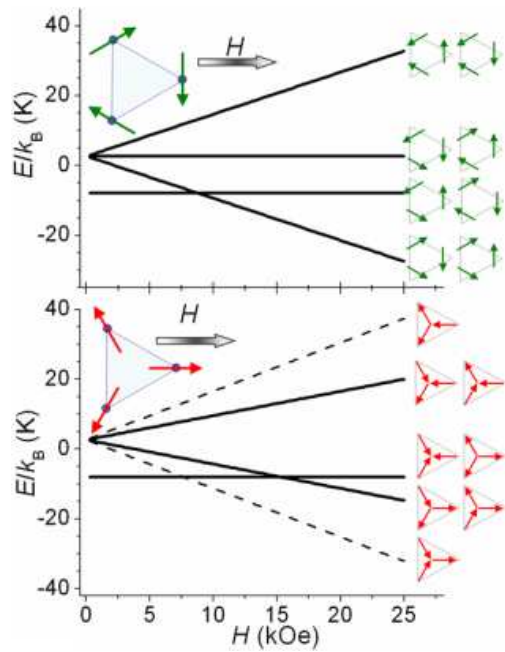}
\end{center}
\caption{(color online) a) Zeeman splitting of the levels due to the application of
the field along the $X$ axis (bisector) calculated with the Spin Hamiltonian (1) and the
parameters indicated in the text, assuming $\phi$ = 90$^{\circ}$. b) Same diagram
calculated with $\phi$ = 0$^{\circ}$. The spin structure for each state is schematized
by the arrows. The non-degenerate states are highlighted by dash lines.
} \label{fig:energies}
\end{figure}

It is interesting to focus on the energy of the eight eigenstates of the Hamiltonian (1)
as a function of the applied magnetic field in the Dy$_3$ plane, as shown in Figure
\ref{fig:energies}. At zero field two states with a zero net magnetic moment, are
degenerate with energy equal to $-3j_{zz}/4$, whereas the other 6 states are also
degenerate with energy equal to $j_{zz}/4$. In an applied magnetic field the energy of
the last six levels depends on the angle the magnetic field forms with the local easy $z$
axes. Let us suppose that the field is applied along one bisector of the triangle as
shown in Figure \ref{fig:energies}. Two limiting scenarios can be observed depending if
the local easy axes are along the edges of the triangle (Fig. \ref{fig:energies}a) or along the bisectors
(Fig. \ref{fig:energies}b).  A jump to saturation magnetization at the first level crossing, $H_1$, is expected in both cases, as indeed observed in Dy$_3$. In general, thanks to the structural non-collinearity of the easy axes, the two states of the ground doublet have opposite vortex chirality, with clockwise or anti-clockwise rotation of the spins.

This simple model cannot however reproduce the weak linear increase in the magnetization
above the step, visible in the inset of Figure \ref{fig:squid}a. We have therefore
formulated a new Hamiltonian model by considering for each Dy$^{3+}$ that the ground state doublet is $|J=15/2, m_J=\pm15/2\rangle$ and that the first excited doublet is$|J=15/2, m_J=\pm13/2\rangle$. These states can be efficiently admixed by the
$J_{+}$ and $J_{-}$ operators of the Zeeman interaction. To avoid over-parameterization
we have limited the treatment to the first excited doublet. The expression for the new
Hamiltonian is:

\begin{eqnarray}
&H=-j \sum^{i,k=1,2,3;i>k}_{\alpha_i, \alpha_k} \cos
(\hat{\alpha_i},\hat{\alpha_k})-g\mu_B \sum^{i}_{\alpha_i}H_{\alpha_i}S_{\alpha_i}
\nonumber \\
&+\frac{\delta}{14}\sum_{i}((\frac{15}{2})^{2}-S_{z_i}^2)
\label{hamilton2}
\end{eqnarray}

where $\alpha_i$ runs over the local axes of the $i$ Dy ion ($\alpha_i$ = $x_i$, $y_i$,
$z_i$). The first term of (2) comes from considering an isotropic (Heisenberg) exchange
between two Dy ions, where $(\hat{\alpha_i},\hat{\alpha_k})$  is the
angle between the $\alpha_i$ and $\alpha_k$  local axes. The last term describes
the single-ion anisotropy and $\delta$ is the zero field splitting between $|J=15/2,
m_J=\pm15/2\rangle$ and $|J=15/2, m_J=\pm13/2\rangle$ states of each Dy$^{3+}$ ion.
The best fit, again performed on the powder data, provided $j =-0.092(2) K$, $g = 1.35(1)$
and $\delta = 102(5) K$. The $g$ value is now close to $4/3$ as expected for $J=15/2$. The
large separation between the two Kramers doublets is in good agreement with what is
reported in the literature for Dy$^{3+}$\cite{Ishi03b,Benelli02}. To reproduce the single crystal data the angle $\phi$
between the $X$ axis and the local $z$ anisotropy was allowed to vary freely. The best
simulation is obtained with the angles $\phi= \pm 17(1)^{\circ} \pm n 60^{\circ}$
(n=0,1,2 ..), where the periodicity results from the symmetry of the cluster.  We have also evaluated the intra-molecular dipolar
contribution to $ j$ as a function of $\phi$, and in the most favourable configuration ($\phi= 90^{\circ} $) it can only account for about half of the observed value.

\begin{figure}[!b]
\begin{center}
       \includegraphics[width=0.35\textwidth]{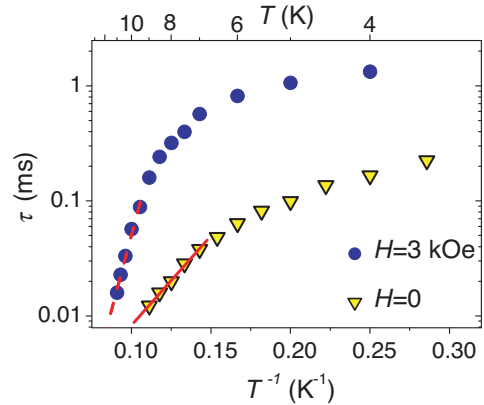}
\end{center}
\caption{(color online) Arrhenius plot of the temperature dependence of the relaxation
time obtained from the frequency dependence of the $ac$ magnetic susceptibility in zero
and applied static field far from a level crossing. The lines represent the linear fit
according to the Arrhenius law of the highest temperature data.  } \label{fig:tau}
\end{figure}

	Susceptibility measurements using standard induction coils in alternating magnetic field allowed us to estimate the
relaxation rate from the frequency dependence of the imaginary component of the susceptibility $\chi^{\prime\prime}$ assuming that, according to
the Debye model, at the maximum of $\chi^{\prime\prime}$ vs $\omega$ curve the
simple relation $\tau=1/\omega$ is valid (see EPAPS for more details)\cite{EPAPS}. When the \emph{ac} field is applied in the
plane of the triangle the relaxation rate decreases on lowering the temperature
following an Arrhenius law, $\tau=\tau_0 \exp (\Delta/k_BT)$, with $\tau_0$=2.5(5)x10$^{-7}$ and  $\Delta$ =36(2) K. At temperatures below \emph{ca.}
7 K the relaxation increases less rapidly, as shown in Figure  \ref{fig:tau}.

The Arrhenius behavior of the relaxation time characterizes a class of molecular materials called
Single Molecule Magnets, SMMs\cite{Eppley97}, where the easy axis magnetic
anisotropy generates a barrier for the reversal of the magnetization giving rise at low
temperature to magnetic bistability and memory effect of pure molecular origin\cite{GatteBook,Christou05,sessoli93}. Dy$_3$,
however, represents the first example where such a slowing down of the magnetization
dynamics occurs even if the overall magnetization lies in an almost isotropic plane
rather than along an easy axis. That is because the system can be better schematized by
three interacting SMMs. In the past pairs of weakly coupled Mn$_4$ clusters have been widely
investigated\cite{Werns02,Hill03} and demonstrated to show quantum
coherence\cite{Tiron03}. However a fascinating new situation is observed in  Dy$_3$ because the two states of the ground doublet are distinguished by a different spin chirality and they cannot be related one to each other by a simple exchange of the magnetic sites. The last is verified even in the special non-chiral case of $\phi=0^{\circ}$.

\begin{figure}[!b]
\begin{center}
       \includegraphics[width=0.40\textwidth]{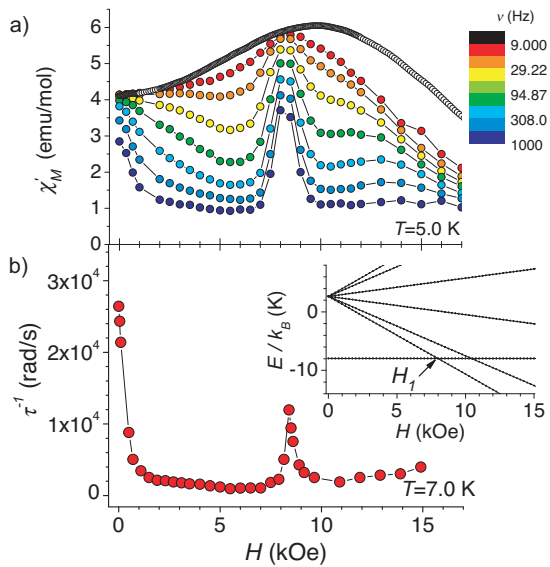}
\end{center}
\caption{(color online) a) Field dependence of the real component at $5 K$ of the $ac$
susceptibility measured with the field along $X$ at logarithmic spaced frequencies.
The empty circles represent the experimental static susceptibility obtained by
derivation of the experimental magnetization curve. b) Field dependence of the
relaxation time measured at $T$=7 K with the field applied along $Y$. In the inset the
Zeeman splitting calculated for $H$ parallel to $Y$ with the best fit parameters discussed
in the text.} \label{fig:susc}
\end{figure}

One of the striking aspect of SMMs is that relaxation at low temperature can occur
through an underbarrier mechanism\cite{GatteBook}. The tunneling is particularly
efficient close to zero field, where the maximum degeneracy of the states is
observed\cite{Friedman96,Thomas96}. The application of a static field has indeed a
strong influence on the dynamics of the magnetization of Dy$_3$, as shown in Fig.
\ref{fig:susc}a  where the real component of the susceptibility, $\chi^{\prime}$, for different frequencies in 1-1000 Hz range is shown as a
function of the field. Around zero field and the first level crossing, $H_1$, the
dynamic susceptibility approaches the equilibrium value, i.e. that obtained from the
derivative of the static magnetization curve recorded at the same temperature, while for
intermediate fields a strong frequency dependent reduction, arising from the
impossibility to follow the oscillating field, is observed. Similar results have been
obtained along $Y$, as shown in Fig  \ref{fig:susc}b where the field dependence of
the relaxation rate at $T$=7 K for this orientation is given. The acceleration of the
relaxation occurs at a field slightly smaller than that corresponding to the maximum of
the static susceptibility, suggesting that only the first crossing is relevant for the
dynamics (inset of Figure \ref{fig:susc}b). Far from the level
crossings, i. e. at $H$=3 kOe,  the relaxation times, given in Fig  \ref{fig:tau}, show a
significant increase of the barrier to  $\Delta$=120(4) K.

The present results are well rationalized by the model we have previously developed.
The response in the \emph{ac} field mainly involves transfer of the population from/to
the ground doublet to/from the first magnetic excited state, implying the reversal of
one spin inside the triangle. Far from any level crossings this seems to occur through
an Orbach process involving the first excited Kramers doublet of each of the Dy$^{3+}$
ions, and its energy gap $\delta$, estimated from the static properties, is found indeed
to be in good agreement with the observed activation energy $\Delta$. On the contrary,
at $H$=0 the reduction of the barrier suggests that an alternative mechanism takes place.
Tunnelling of the magnetization has already been observed for other lanthanide-based SMM
and is attributed to the admixture in zero external field of the ground Kramers doublet
states, made possible by the hyperfine interaction \cite{Giraud01,Ishi05}. At the first level crossing the situation of zero local field is reestablished for a Dy$^{3+}$ ion resulting in the observed fast dynamics.

	To conclude, Dy$_3$ has revealed to be a benchmark system to investigate  non-collinearity in Ising systems. It is worth stressing that non-collinearity is a key feature of molecular magnetism, where complex building-blocks characterized by low symmetry are assembled in more symmetric architectures. Dy$_3$ combines the slow
dynamics of Single Molecule Magnets and the level crossing observed in antiferromagnetic
rings\cite{ taft94,Normand01}. Both types of systems are currently being
investigated for their potential application in quantum
computation\cite{Leuen01,Meier03,Troiani05,Bertaina07} and systems with a
non-magnetic nature of the ground doublet state could be used in order to reduce
decoherence effects due to the fluctuation of local magnetic fields. In the ideal case when the spins lie exactly on the plane of the triangle the dynamics involving the ground doublet is however not directly accessible with magnetometry and requires to be further investigated with more sophisticated techniques, for instance using local probes like muons or neutrons.

We acknowledge financial support from the NE-MAGMANET (FP6-NMP3-CT-2005-515767) and the
German DFG (SPP1137 and the Center for Functional Nanostructures, CFN). W.
Wernsdorfer, A. Vindigni, D. Gatteschi, and M. G. Pini are gratefully
acknowledged for stimulating discussion and helpful suggestions.	

\bibliographystyle{prsty}


\end{document}